\documentclass[10pt,letterpaper]{article}
\usepackage[top=0.85in,left=2.75in,footskip=0.75in]{geometry}

\usepackage{amsmath,amssymb}

\usepackage{changepage}

\usepackage[utf8x]{inputenc}

\usepackage{textcomp,marvosym}

\usepackage{cite}

\usepackage{nameref,hyperref}

\usepackage{microtype}
\DisableLigatures[f]{encoding = *, family = * }

\usepackage[table]{xcolor}

\usepackage{array}

\usepackage{textcomp}

\newcolumntype{+}{!{\vrule width 2pt}}

\newlength\savedwidth

\usepackage[aboveskip=1pt,labelfont=bf,labelsep=period,justification=raggedright,singlelinecheck=off]{caption}

\makeatletter
\renewcommand{\@biblabel}[1]{\quad#1.}
\makeatother

\usepackage{lastpage,fancyhdr,graphicx}
\usepackage{epstopdf}
\fancyheadoffset[L]{2.25in}
\fancyfootoffset[L]{2.25in}
\usepackage{xcolor}

\begin{document}
\vspace*{0.2in}

\begin{flushleft}
{\Large
\textbf\newline{Ten simple rules for training scientists to make better software} 
}
\newline
\\
Kit Gallagher\textsuperscript{1*},
Richard Creswell\textsuperscript{2},
Ben Lambert\textsuperscript{3},
Martin Robinson\textsuperscript{2},
Chon Lok Lei\textsuperscript{4},
Gary R. Mirams\textsuperscript{5},
David J. Gavaghan\textsuperscript{1*}
\\
\bigskip
\textbf{1} Doctoral Training Centre, University of Oxford, Oxford, UK
\\
\textbf{2} Department of Computer Science, University of Oxford, Oxford, UK
\\
\textbf{3} Department of Statistics, University of Oxford, Oxford, UK
\\
\textbf{4} Department of Biomedical Sciences, Faculty of Health Sciences, University of Macau, Macau, China
\\
\textbf{5} Centre for Mathematical Medicine \& Biology, School of Mathematical Sciences, University of Nottingham, Nottingham, UK
\\
\bigskip

* gallagher@maths.ox.ac.uk; david.gavaghan@dtc.ox.ac.uk

\end{flushleft}



\section*{Introduction}



Computational methods and associated software implementations are central to every field of scientific investigation. 
Modern biological research relies heavily on the development of software tools to process and organize increasingly large datasets, simulate complex mechanistic models, provide tools for the analysis and management of data, and visualize and organize outputs\cite{Ghosh2011, Markowetz2017, Baxter2006}. Such software varies widely in its scope, complexity and potential for re-use, from single-use analysis scripts that accompany journal publications to domain-specific packages (such as molecular dynamics simulators \cite{oxDNA, LAMMPS}), common numerical methods (such as finite element methods \cite{FiPy2009} or optimization algorithms \cite{Gad2023}), and finally fundamental scientific software (such the numerical methods package `numpy' \cite{Harris2020array}).

For valid usage in research, it is essential that this software is both openly available and accurately implements its intended functionality. Accessibility of code has improved significantly in recent years \cite{Cadwallader2022}, and it is increasingly accepted that research papers should be accompanied by accurate code scripts, which are subject to peer review alongside the other methods of the research. However, this has simultaneously highlighted the role of computational science in the so-called `reproducibility crisis' \cite{Baker2016}, where multiple cross-disciplinary meta-analyses have indicated that less than half of published code may be run without errors \cite{Trisovic2022, Konkol2018, CChang2022, Stodden2018}, and as little as 5\% can replicate the primary results of the associated paper \cite{Pimentel2019}.

This causes a multitude of negative effects on scientific research including a lack of transparency and open access \cite{Howison2015}, poor development and deployment practices \cite{Katz2014}, and a lack of executable reproducibility \textemdash\ where code cannot even be run \cite{Strijkers2011}.
This also undermines the productivity of the research software base, as any researchers wishing to use the same computational framework are then forced to re-implement this in their own software.

Beyond basic reproducibility, higher-quality software possesses additional qualities such as \emph{extensibility}, \emph{reliability} and \emph{reusability.} These characteristics arise from carefully designed, well-documented, and appropriately maintained code, and they enable research software to more thoroughly and efficiently support scientific progress \textemdash\ for example, by allowing a software package developed by one research group to be picked up by another which goes on to add additional features to address further scientific questions. The term \emph{sustainable} (not to be confused with environmentally friendly software) has been adopted to refer to software that is reliable, reproducible, and reusable \cite{Crouch2013}. 

Given the importance of high-quality software to effective research in computational biology, there has been significant literature on ensuring reproducibility \cite{Sandve2013, Peng2011} and good development practices \cite{Osborne2014,Brack2022, List2017} in computational research. 
Indeed, several other Ten Simple Rules articles have already provided excellent descriptions of best software development practices to aspire towards, and we point the reader to these guides on documentation \cite{Lee2018}, usability \cite{List2017}, robustness \cite{Taschuk2017} and version control \cite{PerezRiverol2016}. 

However, less attention has been devoted to specific teaching strategies which are effective at nurturing in researchers the complex skillset required to produce high-quality software that underpins both academic and industrial biomedical research. 
Biologists and computational researchers, even if aware of the importance of high-quality software to their research, are typically left to fend for themselves in developing the necessary skills to produce reusable software.
Although training resources are available (for example, courses offered by the Software Sustainability Institute to UK-based researchers), many doctoral training programmes overlook extensive formal education in effective software engineering.

Two recent articles in the Ten Simple Rules collection \cite{Carey2018, Reyes2022} have discussed the teaching of foundational computer science and coding techniques to biology students. We advance this discussion by describing the specific steps for effectively teaching the necessary skills a scientist needs to develop sustainable software packages that are fit for (re-)use in academic research or more widely. We advocate that future researchers receive extended training in software engineering, moving beyond few-day training sessions and forming a substantial and integrated portion of their scientific education. Although our advice is likely to be applicable to all students and researchers hoping to improve their software development skills, our guidelines are directed towards an audience of students who have some programming literacy but little formal training in software engineering, typical of early doctoral students. These practices are also applicable outside of doctoral training environments, and we believe they should form a key part of postgraduate training schemes more generally in the life sciences.

The following rules have been fine-tuned through generations of doctoral students at the \textit{EPSRC CDT in Sustainable Approaches to Biomedical Science: Responsible and Reproducible Research (SABS:R\textsuperscript{3}) CDT} at the University of Oxford. The SABS:R\textsuperscript{3} program trains doctoral students in cutting-edge, collaborative systems approaches to biomedical research, with a strong focus on computational methods. Perhaps uniquely, it provides both comprehensive training in advanced software development and software engineering to all of its students, and introduces them to state-of-the-art techniques applicable to industrially-derived research within the biomedical sciences. 

Students initially take a three-week classroom training program in the principles of software engineering. This course is designed to introduce important themes in software development and convey their importance; we discuss key aspects of this in the first four rules. National and international consortia can provide resources and expertise to support institutions in running these programs - the SABS:R\textsuperscript{3} classroom section was supplemented by training from the Software Sustainability Institute.

To make this training immediately relevant, all students subsequently undertake an industry-supported group software development project over their first year, allowing them to learn and apply their software skills in a realistic setting. These projects require students to develop high-quality software in support of some scientific or industrial project or investigation, involving a variety of fields including epidemiology, pharmacokinetics, and medical imaging. Two such projects have additionally led to scientific publications \cite{Creswell2022, Gallagher2024}. We have fine-tuned the design and implementation of these projects over a series of student cohorts and share some of the lessons we have learned in Rules 5--8, before finally discussing the output from these projects, and how they may be sustained after the students finish their training.

\section*{Rule 1: Emphasize the value of good software development}

It has widely been accepted within industrial software development communities (and is increasingly accepted within academic environments) that investing additional time to uphold common software engineering practices improves overall productivity across a wide range of project scales and fields \cite{Blackburn1996, Canedo2019, chue_hong_2023_8205595}. 

However, this translation of this acceptance to academia has been limited by insufficient prior training, as mentioned in the introduction, as well as a lack of its perceived value and importance. Within the results-focused setting of academia, good software development has often been considered a secondary priority \textemdash\ `nice if you have time, but not essential'. This may be attributed to the perceived lack of academic recognition for software outputs \cite{Way2021, Hafer2009}, diminishing the motivation and financial support for upholding good practices. This is exacerbated in the time-pressured environment of a PhD, where the limited time frame encourages a focus on obtaining immediate results over longer-term investments, such as a high-quality and reusable code base. 

\begin{figure}[htb]
    \centering
    \includegraphics[width=0.6\linewidth]{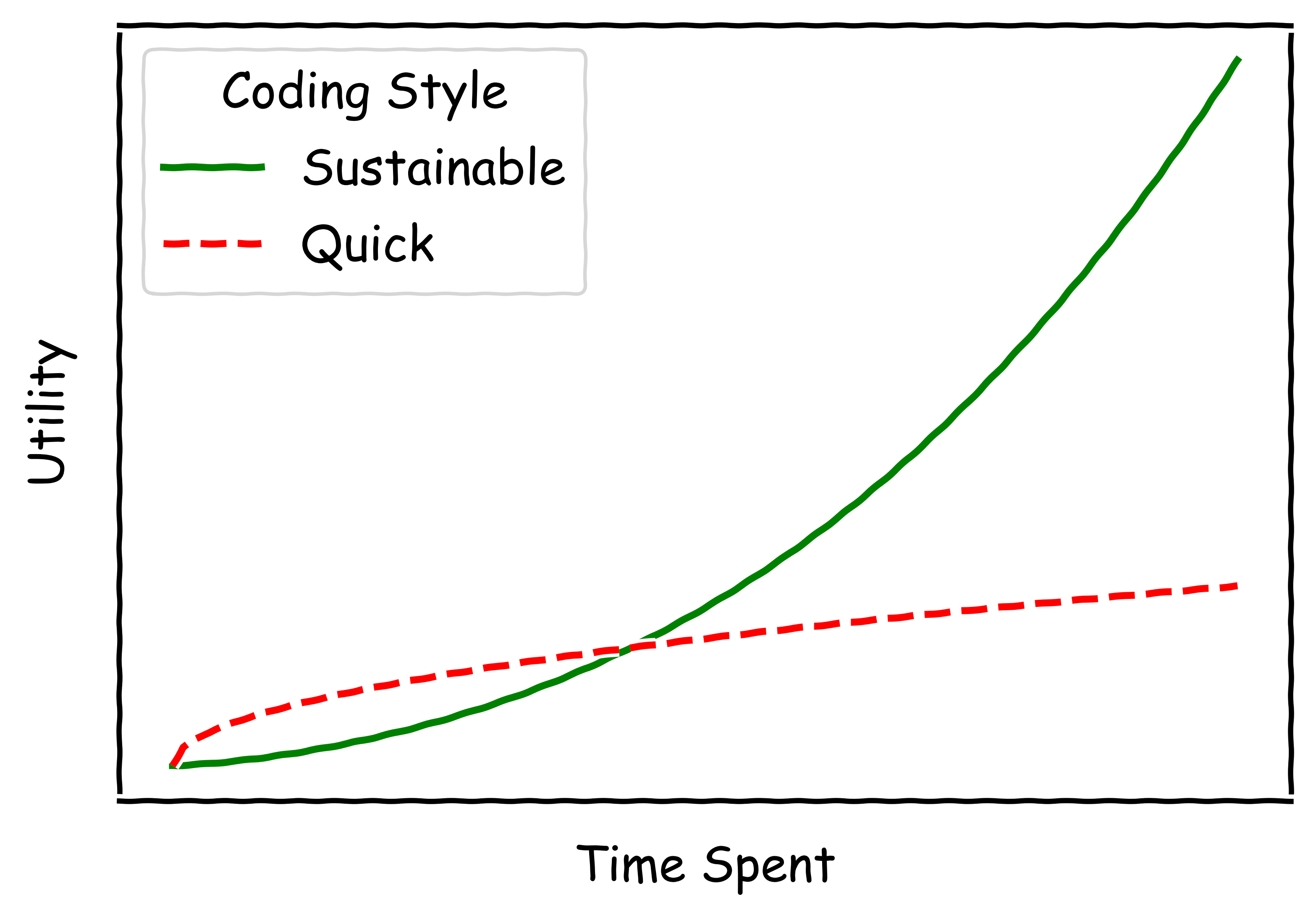}
    \caption{While sustainable approaches to software development can require more effort than a quick-and-dirty coding style initially, they quickly become time-saving over medium-to-long time scales.}
    \label{fig:sustainable_vs_quick}
\end{figure}

Much research software is initially developed for highly specialized applications (for example, a pipeline of idiosyncratic procedures to process a particular dataset). In this context, reusability and extensibility may not seem essential at the outset of the software's lifespan, but their utility may become apparent later on (e.g., several years later, if a new PhD student is tasked with processing a new dataset with similar characteristics, are they able to draw upon the previous pipeline, or must they start over from scratch?). This increased productivity quickly pays back the initial time investment required, and allows the easy integration of small pipelines into large and varied code bases, increasing their potential application (Fig \ref{fig:sustainable_vs_quick}),  While this paper focuses on the development of large-scale software packages intended for immediate reuse; many of the skills we aim to inculcate in researchers will equally well serve them in the development of higher-quality analysis scripts.

By emphasizing the advantages of good software development with students at the start of their research careers, it is possible to dispel any preconceived assumptions that careful coding practices will inherently slow down the pace of research. Our experience, both as PhD supervisors and PhD students, is that while students are often enthusiastic to learn software development as a new skill, the benefits to scientific research of a formal grounding in sound software engineering practices may not be immediately apparent at the start of the PhD. The advantage of, for example, taking a test-driven approach is typically realized later in the PhD, as codes developed early in the PhD can be extended and built upon with confidence. Ultimately, good software practices such as automated testing and prudent refactoring can prevent bugs, flawed results and retractions \cite{Miller2006}. 

As we discuss further in Rule 8 below, these benefits can be reinforced through structured peer mentoring. Over time, this results in a body of robust and reusable open-source software underpinning the long-term research within the group, and these benefits become more immediately obvious and self-sustaining.

\section*{Rule 2: Support students to develop good coding practices}

There are a variety of schools of thought for proper coding practices, and giving students a degree of freedom over the standards they adopt can encourage accountability and teach the role of such practices. However, it is essential to provide students with direction for what good code is. This will help student groups to come to a consensus that aligns with widely used programming practices, while still leaving enough room for interpretation. It is often helpful to introduce students to general standards used by existing projects/domains, before allowing them to adjust it based on the specific requirements of their work. This is a key concept in the FAIR (Findable, Accessible, Interoperable and Reusable) principles for research software \cite{Barker2022}, where accordance with domain-specific standards supports code reuse by ensuring their code can interoperate with other codebases in the future.

When introducing coding standards, and encouraging students to think critically about how their code should be written, it can be helpful to direct them to think about points such as the following:

\begin{itemize}    
    \item Can you choose meaningful and understandable variable and function names, so that your code is somewhat self-documenting?
    \item What are the naming conventions in your programming language? (For example, in Python, the PEP 8 Style Guide prescribes the use of snake case for function and variable names, and camel case for class names.)
    \item How will you lay out your code? While aspects such as indentation are part of the syntax in some languages (like Python), this (along with the use of whitespace and allowed line length) is a stylistic choice in others.
    \item How will you document your code, and communicate changes (such as refactoring) to others working on the same codebase?
\end{itemize}

\begin{figure}[ht]
    \centering
    \includegraphics[width=.8\textwidth]{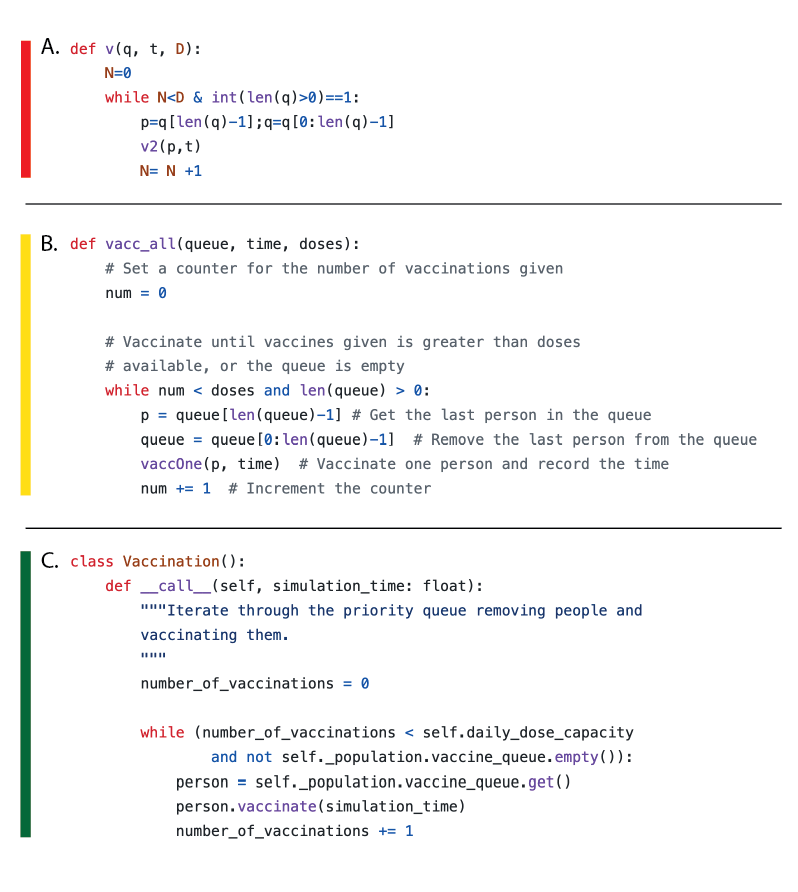}
    \caption{Codes snippets A., B., and C.\ are all valid Python and implement identical functionality. However, they vary in readability: A.\ is difficult to understand for anyone except the original author; B.\ is much improved, but perhaps suffers in readability due to verbose commenting; C., however, should be directly interpretable on account of meaningful variable names and sensible object-oriented architecture.}
    \label{fig:good_and_bad_code}
\end{figure}

The impact of such changes on the readability of the code (without changing its function) can be seen in Fig \ref{fig:good_and_bad_code}. 
Similarly, it can be helpful to warn students of so-called `code smells', which suggest failures to adhere to good software practices and are associated with increased code maintenance \cite{Yamashita2012, Lacerda2020}. Students should be asked to avoid situations such as the following (unless they have a good reason to break the guidelines):

\begin{itemize}
    \item Functions should not contain more than 20 lines.
    \item Repeated code should be combined into a function for re-use.
    \item Unused features (created to anticipate future functionality that never gets implemented) should be removed.
    \item Standard library functions/data types (i.e. min/max functions or arrays) should not be reimplemented from scratch.
    \item Long parameter lists (i.e. with more than four parameters) may be collapsed into a single object.
    \item Functions/classes/methods should not have multiple responsibilities. If they do, their scope should be simplified by defining additional functions/classes/methods.
\end{itemize}

While these approaches may require some refactoring of the student's code, this typically results in significant time savings over the lifetime of the code, reducing the amount of time spent debugging \textemdash\ the balance of such approaches is discussed by Balaban et al. \cite{Balaban2021}.
These good practices may be developed interactively through exercises identifying antipattern in deliberately poorly written code, and then rectifying the issues with it. This can also help students appreciate the difficulties associated with working on poorly written code, and motivate the importance of good coding practices before they begin working on larger-scale projects. Additionally, students should take the time to explore the features of static analysis tools and Integrated Development Environments to support their efforts to follow good coding practices (see Rule 9).

\section*{Rule 3: Advocate the role and importance of open-source and collaborative development}
\begin{figure}[ht]
    \centering
    \includegraphics[width=.45\textwidth]{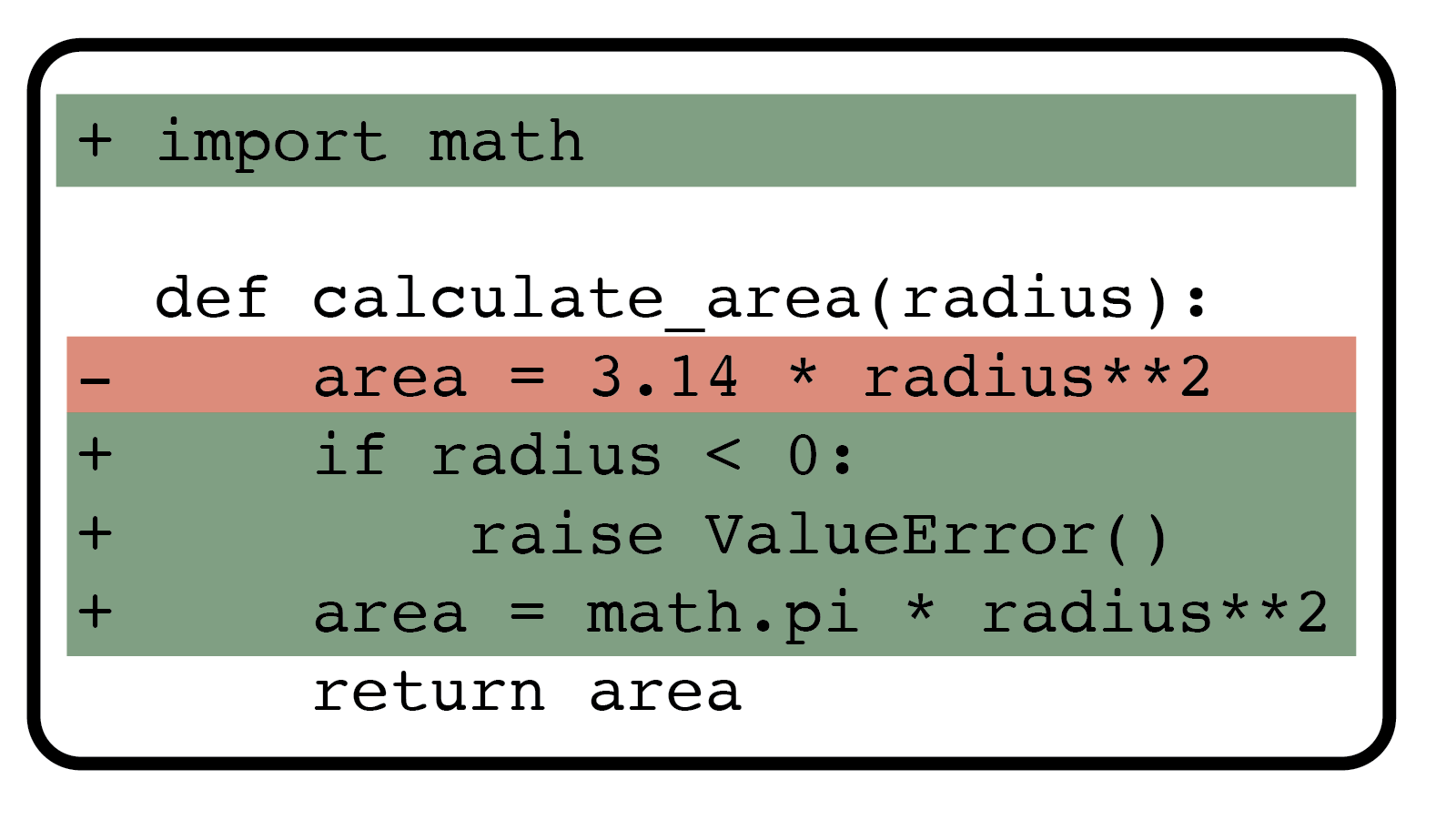}
    \caption{Version control tools such as GitHub make it easy for developers to keep track of changes in a software codebase.}
    \label{fig:rule3}
\end{figure}

We encourage open-source development using sites such as GitHub (Fig \ref{fig:rule3}), and find that most students are enthusiastic about sharing their work in this way. The benefits of open-source development for early career researchers \cite{Allen2019, McKiernan2016}, software quality \cite{Fitzgerald2006, Paulson2004}, and research in general \cite{vonKrogh2007, Pearce2012} are well documented.

This also prompts students to think about how their work may be shared, and we specifically teach the students how to package and distribute their code, including managing dependencies and publishing to package management systems such as pip or CRAN. 
The students also explore how they can facilitate other researchers' use of their code, such as by providing online documentation, installation instructions/executable files and example workflows. 
Finally, we cover code maintenance, so that students are familiar with the complete life-cycle of a software product, and can make informed decisions on maintaining or withdrawing support for different operating systems or versions of external dependencies.

We recognize that open source development may not always be completely possible, for example, if industrially-partnered research projects involve proprietary data or methods that must be hidden, but in these situations believe it is helpful to train students to share their code to the extent possible. Collaborative work is still possible in this context, through the careful use of private repositories, or excluding sensitive data from public versions of the repository. 
Students are also taught about the importance of software licensing to allow the reuse of their software, and the different approaches that licenses take to redistribution and proprietization. Clear licensing and accessible software (including metadata) are both key features of the FAIR principles for research software \cite{Barker2022}.

Having a public-facing repository also means that software may attract those in the wider community to act as co-developers, which can also encourage students to take more ownership and care of the quality of the repository. As other users can see, use or even contribute to their code, software development principles discussed in Rule 2 such as ensuring code quality and good documentation have immediate benefits. 
The potential for wider involvement also prompts students to review how they might support third parties wishing to contribute, seek support or report issues in the software. As part of this, students are guided to promote recommended community standards by providing public contribution guidelines, issue templates, discussion boards and security policies for their software \cite{Nakakoji2002}.
Previous teaching projects \cite{Epiabm2022} within the SABS:R\textsuperscript{3} program have attracted external collaborators across the globe, as a result of effective publishing through GitHub and clear open-source contribution guidelines.

\section*{Rule 4: Use agile development practices}

Among different development paradigms, so-called `agile' development has many features that are beneficial to scientific research. Project goals are typically not fully defined at the conception of a project and are often informed by preliminary and intermediate results, so the ability to easily refactor the codebase is essential. 
Agile development methodologies account for this through concise planning horizons (often termed `sprints' or `iterations'), working towards coding targets that can be feasibly accomplished in a reasonable time frame (typically one month). 
This approach facilitates the rapid creation of working prototypes and mitigates the phenomenon of `analysis paralysis', a pitfall that arises when attempting to anticipate too many potential future requirements.

Whilst there are several frameworks for agile development such as `Scrum' or `eXtreme Programming' \cite{Beck2004} we emphasize the basic principles in our courses rather than any particular approach. 
The key principles are continuous integration (that is, always having a codebase that passes automated tests and is safe to deploy, even if lacking features); test-driven development (wherein tests are written before the source code itself, requiring students to consider the best interface for new code), short-range planning; and frequent code refactoring (reorganization and tidying). 
These characteristics provide a good framework to adapt to changing demands and evolving research questions; importantly, they produce software that is well tested \textemdash\ whose outputs do not change without developers realizing it \cite{Ahmed2010, Sletholt2011, Madeyski2007}.

\section*{Rule 5: Create a real-world project that is ambitious, relevant, and exciting}
In general, we learn skill-based knowledge best by doing, rather than through taught lectures or textbooks \cite{Freeman2014, Prince2004}. For this reason, we developed annual group projects for students to consolidate the software development skills introduced in taught courses. These problems should not be toy projects; they should be challenging, real-world problems that have not been tackled before. We have found that this novelty motivates students to engage more directly with the project \cite{Ros2010}. 

\begin{figure}[ht]
    \centering
    \includegraphics[width=.45\textwidth]{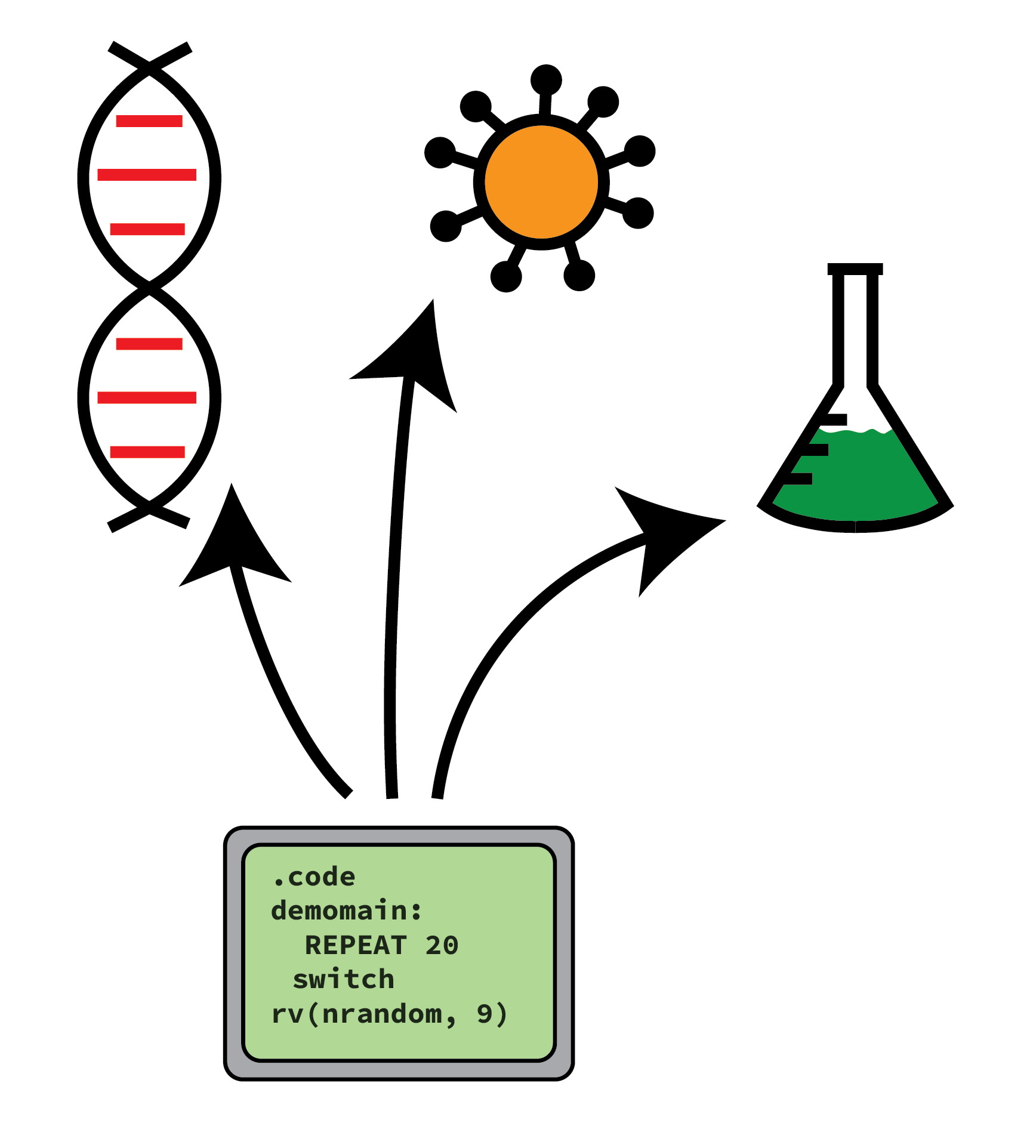}
    \caption{To keep students of all interests and backgrounds excited, we recommend software projects that are oriented towards appealing scientific subjects (e.g., biology, epidemiology, chemistry), rather than projects that dwell merely on more foundational computer science or algorithmic tasks.}
    \label{fig:rule5}
\end{figure}

These projects should also have wider relevance outside of pure pedagogy, be it to academic research or industrial applications (Fig \ref{fig:rule5}). This typically necessitates a larger scale problem than typical teaching/textbook problems, requiring a team of students to tackle it together. This resembles effective software engineering in industry, where good software is developed in large, multi-disciplinary groups. This will be in contrast with the primarily individual experience that students typically gain during undergraduate research projects, where individual scripts are written to answer a specific research question rather than creating a tool for exploring a wider field of research. These larger projects help to contextualize the role and uses of research software, and develop students' experience and confidence in working with a large codebase. A few examples of projects we have previously run with student cohorts are linked below, to give a perspective of both the wide range of research areas covered and the scope of possible outcomes of these projects:

\begin{itemize}
    \item \href{https://github.com/SABS-R3-Epidemiology/epiabm}{Epidemiological agent-based modeling (EpiABM) software, based on CovidSim model developed at Imperial College London \cite{Epiabm2022}}
    \item \href{https://github.com/SABS-Group-2-2021-22/drug-discovery-game-app}{An educational game based on the process of designing new drugs, developed in collaboration with Roche \cite{DrugDiscoveryApp}}
    \item \href{https://github.com/Extensible-Clinical-Imaging-QC-Tool/ECIQC}{Extensible quality control tool for clinical images, developed in collaboration with GE Healthcare \cite{ECIQT}}
\end{itemize}

Ensuring that these projects have tangible outputs is also key to student engagement. The key output for these projects is a complete software package, which should be open-source if possible (see Rule 3) to enable the wider research community to engage with this work. Additionally, because the projects are drawn from real-world research questions, successful software outputs will involve insightful, novel results shedding light on some scientific problem. This gives students the motivation to apply rigorous scientific thinking to any modeling assumptions inherent in their software, and to examine their software's behavior in some rigorous way (for example, by comparing its predictions to existing benchmark data or established comparator methods).

Furthermore, in SABS:R\textsuperscript{3}, projects were developed in collaboration with industrial partners to tackle problems in their respective fields, and many of these industrial partners have subsequently used the open-source research outputs in-house. The industrial partners have also supported the projects, funding access to compute resources or providing access to databases to enable these projects to have real-world impact. The impact may also be tracked more broadly, through a combination of direct software engagement (i.e. downloads or citations), or public event engagement (such as the attendance at a software workshop). Development of web-hosted user-friendly interfaces allows greater quantification of the impact through telemetry tracking the number of users, use frequency, and typical use cases.

Providing publishing opportunities, either within dedicated software journals or through scientific journals in cases where the software has been used for notable scientific applications, can also provide a concrete end goal for students. Thus far, three cohorts within the SABS:R\textsuperscript{3} program have published (or prepared for publication) descriptions of their software and associated scientific findings in reputable journals \cite{Creswell2022, Gallagher2024}; further scientific and software publications are anticipated in future years of the program.

\section*{Rule 6: Set up long-term projects for multiple cohorts of students}

\begin{figure}[ht]
    \centering
    \includegraphics[width=\linewidth]{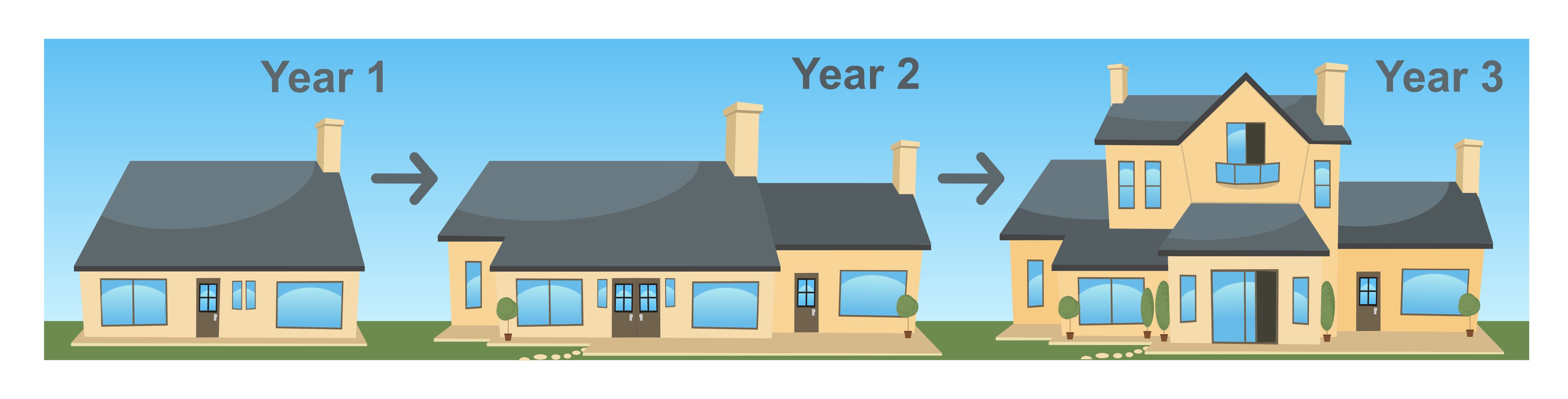}
    \caption{While all year-long projects should produce a complete and usable product, these can then be further extended by subsequent cohorts. Such extensions should not be trivial `bolt-ons', but rather fundamentally extend the functionality of the software output, which may require students to modify/refactor existing elements of the codebase.}
    \label{fig:multi_cohort}
\end{figure}

We have typically run multi-cohort projects, with a new cohort of students working to extend an existing codebase developed by a previous cohort in the preceding years. The benefits of this are twofold. Firstly it allows more ambitious projects to be tackled which might not be attainable in a year (Fig \ref{fig:multi_cohort}). 
Secondly, developing someone else's codebase is an essential software skill and encourages students to write their code with future developers in mind.

While most teaching in academic settings focuses on developing a new codebase, this is rarely the reality of software development in academia or industry, and it is common for new starters to struggle to adapt to working on legacy code instead of developing new projects \cite{Craig2018}. Providing experience of this in a teaching setting (and often with the luxury of access to the previous developers \textemdash\ see Rule 8) gives students the skills required to implement changes in long-standing research software codebases.

Projects which merely involve the maintenance of an existing codebase, without adding any new features, are in our experience less likely to spur significant student interest, so we recommend that legacy software projects always involve adding some new features. 

Multi-cohort projects can also adapt to changes in the project motivation \textemdash\ for example, a project developing broad epidemiological inference software at the start of the COVID-19 pandemic \cite{Creswell2022}, was redirected to develop agent-based modeling software \cite{Gallagher2024} following the high-profile use of this approach in UK governmental response to the pandemic \cite{Ferguson2020}. This software package was recently extended by a subsequent year group to interface with spatial data repositories and integrate accurate population information into epidemiological models \cite{Herriott2023}. This approach ensured the relevancy of the software being developed, and that it would continue to be supported and maintained after the original developers graduated from the teaching course. To develop skills in software maintenance, we also encourage students (even those not involved in peer supervision (see Rule 8)) to revisit any maintenance-related concerns that arise after they complete the training program.

In addition to this focus on developing pre-existing codebases, we recognize that the process of embarking on a `greenfield' software project \textemdash\ where students must create a new software library from scratch \textemdash\ still provides valuable lessons in how to set up the fundamental architecture and collaborative infrastructure of a software project. We have therefore found it valuable to set aside the first few days of the training for a greenfield `mini-project'. This approach also allows instructors to gain a detailed insight into student ability levels, strengths, and weaknesses, which can guide the goals and organization of the primary project.

\section*{Rule 7: Encourage collaboration within mixed-ability groups}

Software development is rarely a solo endeavor, and this should be reflected in the teaching environment. Group work encourages students to learn from each other's coding styles, as well as `soft skills' such as communication and teamwork within the specific context of software development. 

\begin{figure}[ht]
    \centering
    \includegraphics[width=\textwidth]{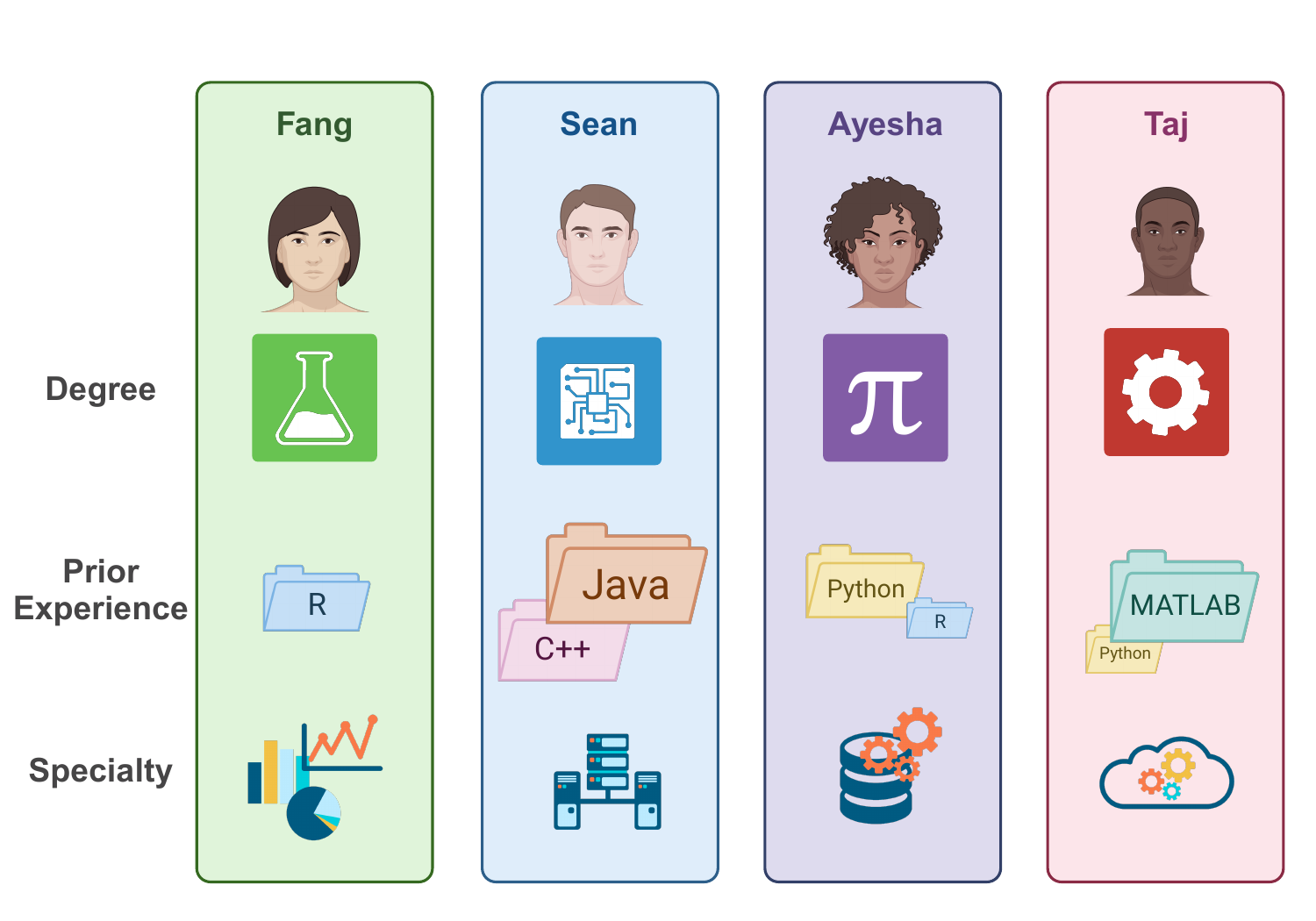}
    \caption{Cohort heterogeneity can take many forms, such as previous academic degrees/professional experience, prior fluency in different programming languages (which may or may not include the teaching language of the course), and familiarity with wider software development skills. Course leaders can take advantage of this range when designing and supervising projects so that all students can both utilize their own strengths and learn skills from others. Created with BioRender.com.}
    \label{fig:cohort_heterogeneity}
\end{figure}

Students are not a homogeneous group however; they will enter any program with differing levels of experience. While this can be seen as a disadvantage, collaboration within groups allows students to share their various individual skill sets, becoming their own teachers as they share their particular specialties \cite{Porter2016}. For example, students with more experience can gravitate towards leadership or design roles, while the less experienced students can learn from more experienced, for e.g. through pair programming \cite{Hanks2011, Salleh2011}. Ideally, groups should be large enough to contain students with a variety of skill levels (Fig \ref{fig:cohort_heterogeneity}) and enable some degree of specialization between students, but small and cohesive enough for all students to maintain familiarity with all aspects of the project. We have found that groups of 4--5 students work very well.

These groups also give students a strong motivation to develop project management skills via planning discussions and delegation of responsibilities among themselves. Although formal training in project management techniques may be available at some institutions or from other sources, such training is typically classroom-based and often optional. This means that students will benefit from consolidating these skills in a practical context, to refresh concepts they may have previously encountered in taught modules. Group software work also requires the use of version control software, and developing skills such as branch management, bug tracking, and code review that students are taught at the start of the course, but may not have applied during solo projects. 

With larger-scale projects (see Rule 5), it is inevitable (and indeed desirable) that projects will be modular, and readily broken into separate tasks. Despite this, it is important to ensure that the different aspects of the project still interact where possible so that learners are not working in isolation on their sections. This interaction often happens explicitly (for example when developing a computational pipeline where elements must interface effectively), but ensuring individual students are not isolated working on separate aspects of the codebase is crucial to ensuring that the students learn from each other's experience. Peer-to-peer learning and assessment may be strengthened by promoting a shared responsibility in code review, developing teamwork and communication skills within a computational environment as well as exposing students to new approaches and design patterns in their peers' code. These approaches also help avoid premature specialization, where students only learn about their own features and miss out on skills from other areas;  similarly, all learners should be comfortable documenting and testing their code, rather than leaving this to a single individual.

We have found it most effective for students to have the chance to work together in person, enabling collaborative activities such as regular stand-up meetings and pair programming. When students are working remotely, appropriate collaboration tools should be employed and project leaders should ensure that all students remain actively involved in the project (e.g., via holding regular group and one-on-one meetings, or monitoring contributions to identify any students who might be withdrawing from the project), for the benefits of collaborative activities such as pair programming to still be realized \cite{cockburn2000costs, schummer2009understanding}. For example, we have found it effective when students engage in pair programming remotely using the screen-sharing facility offered by most video conferencing platforms, possibly in conjunction with the live-coding feature present in some modern IDEs such as VSCode.

\section*{Rule 8: Organize supervision from former students and peers}
Practical courses at a graduate level are often limited in scope by the availability of senior academics to offer supervision. Instead, more senior students on the course can act in supervisory roles for these projects \textemdash\ able to provide hands-on supervision, and a more informal resource (away from senior academics) to answer questions about unfamiliar aspects of software development (such as unit testing). Drawing students from previous cohorts that worked on the same project is particularly advantageous (and is a proven technique in open source development: used, for example, in Google's Summer of Code [\url{https://google.github.io/gsocguides/mentor/}]), as this ensures that student supervisors have a strong familiarity with the underlying codebase. This relationship is also beneficial for the supervising student \textemdash\ as well as learning through their teaching they also have first-hand experience of the struggles new developers may have picking up their codebase, and how they could mitigate this in future projects.

This supervision also mimics professional structures common in industry, where junior developers will typically submit work to senior developers for code review. Another good practice for learners
(and supervisors) who may later wish to transition to careers in the tech sector is the formal use of version control features and code review. Code review can also help to update other members of the team on features within the code, and formalize the protocol for introducing significant changes into the main codebase, including those that affect wider functionality.

Encouraging peer code review, where learners begin to review each other's pull requests, is also highly beneficial to ensure continual feedback on code outputs. This form of comprehensive peer supervision can be supported by continuous integration metrics such as test coverage and code quality, reducing some of the supervision requirements in student projects by automatically highlighting issues in the students' code that they must resolve themselves \cite{Sus2012}.
Peer-to-peer feedback can also be used to derive assessment metrics of student performance \cite{clark2005self}.

\section*{Rule 9: Focus on the process, not the final outcome}
Although projects should be ambitious and exciting (Rule 5), the final outcomes of these projects should never come at the expense of the learning process. Course leaders should endeavor to foster a creative environment where students are focused on achieving the desired functionality in the best way possible (rather than the most functionality possible), enabling students to produce higher quality outputs and develop stronger software development skills to take forward into their future projects.

Students will gain more from having the time and freedom to write higher-quality code at a slower pace, rather than rushing to complete targets.  For this reason, we do not recommend formal assessment (i.e., grades) of student contributions to software training projects; instead, informal and constructive feedback on students' rate of progress, strengths and areas for improvement should be provided by instructors and mentors. Meanwhile, students should also be continuously receiving detailed, practical feedback on their code via code review from their peers or course demonstrators.

This focus on personal development also provides students with a unique opportunity to learn features and patterns of the programming language with which they are unfamiliar, in a way that is rarely achieved elsewhere in their studies when working to deadlines. However, students should avoid using this freedom to delve into obscure, esoteric, or overly academic coding styles, and they should be encouraged to keep the practical goals of the project in mind as they explore more advanced concepts in their programming language (see Rule 2).

If the initial project aims turn out to be too ambitious, projects should be rescoped to focus on some smaller goals that the students can more confidently achieve without ever feeling the need to forego good software engineering practices. Alternatively, students can be divided into different streams working on different aspects of the project, or initially be assigned an easier toy project to develop various software development skills before launching into more ambitious year-long projects where they learn to balance best practices with time and resource limitations. 

Another key benefit of this teaching program is the opportunity to explore and develop familiarity with common software development tools, as proficiency in these will boost their productivity throughout both their PhD and subsequent career choices in computational fields. As introduced in Rule 7, one key aspect of this is familiarity with version control tools such as Git, and the importance of this skill has been addressed in `Ten Simple Rules for Taking Advantage of Git and GitHub' \cite{PerezRiverol2016}. Students should be taught and encouraged to utilize features of GitHub widely, for example, to document bugs and planned features through issues or peer-review each other's code through pull requests. An additional advantage of students engaging with GitHub features is that they will build up a public history of code commits, issues, pull request reviews, and other software-related tasks; such track records are extremely valuable evidence of software-related competency during the hiring process for both industry and academic jobs involving software development.

Similarly, familiarity with the functionality of Integrated Development Environments (IDEs), including built-in debugging and refactoring tools, is important for efficient software development and is associated with students' improved learning and productivity \cite{Dyke2011}. Learners may also find it helpful to use static analysis tools, such as PVS-Studio (for C, C++, and Java) or flake8 (for Python), which provide feedback on code style to develop awareness of clean code practices.

Furthermore, outcomes may not simply be limited to the academic output or even the function of the code. The importance of software development practices can be emphasized, including software development goals as well as functionality \textemdash\ for example, a group may target publishing complete documentation online, or 100\% unit test coverage. When a software project is used as part of a scientific publication, the quality of the underlying software design and engineering (for example, in user-friendliness, extensibility, and reproducibility) should be highlighted in the publication.

\section*{Rule 10: Build a community to `future-proof' your project}

\begin{figure}[ht]
    \centering
    \includegraphics[width=\textwidth]{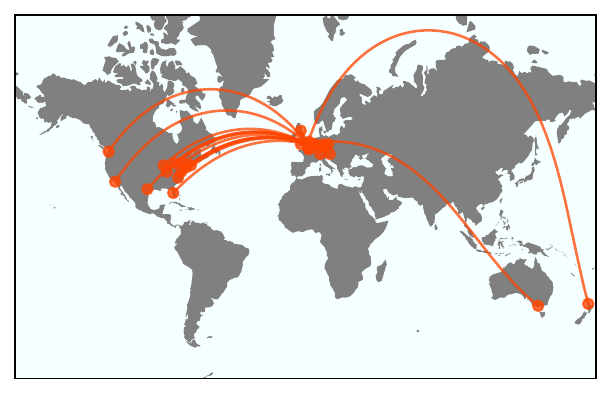}
    \caption{We downloaded the first 50 citing papers of Chaste \cite{Mirams2013} according to Google Scholar (ordered by relevance), and, for each paper, we plot the location of the city of the first corresponding author's institution as an orange dot, with orange lines indicating the great circle path from that location to Oxford, England where Chaste was originally developed. Public domain continents data from Natural Earth.}
    \label{fig:rule10}
\end{figure}

Our experience over the last two decades highlights the potential benefits of providing advanced training in software development in the context of a `live' applied research project. Our first experience with this approach was in 2005, when we used the initial development of the Chaste (Cancer, heart and soft tissue environment) physiological modeling software \cite{Pitt2009, Osborne2010, Mirams2013, cooper2020chaste} to teach test-driven, agile software development practices to a cohort of ten PhD students and several post-doctoral researchers. Over the intervening period, Chaste has become one of the leading such platforms and is in continual use by research groups across the world (Fig \ref{fig:rule10}). Over 50 early career researchers have contributed to its development, and several of the original Chaste developers now use it within their own research groups in academia, industry, and at regulatory authorities.

Although this success is unlikely to be replicated in every such training-driven and hence student-led development project, our similar experience in the development of PINTS (Probabilistic inference for noisy time series) \cite{Clerx2019} suggests that this approach has the potential to deliver sustainable and reusable software platforms to the research community with surprising frequency. The key here, we believe, is the generation of a community of researcher-developers who take collective ownership of the code and, through joint publications and frequent code releases, have a joint interest in its continued maintenance and use.
This community stemmed from an early focus on developing the user base of this software from the original developers to a wider, multi-institutional community. For further details on this development, we refer readers to excellent Ten Simple Rules articles that have previously addressed community-building \cite{Sholler2019}, and promotion/sponsorship \cite{Prli2012}.


Communities of software users, developers, and maintainers can take many forms and may range in scale from local collaborations within a university up to the more global network we observed with Chaste (Fig \ref{fig:rule10}). Although such communities will inevitably tend to develop alongside existing collaborative relationships, it is also possible to start building connections with developers in other regions of the world using online, collaborative development tools such as GitHub (see Rule 3).

Building such communities can future-proof a code base, and perhaps suggest a new and more sustainable model for the development of the computational tools and software that increasingly underpin research.

\section*{Conclusion}
Instilling good software development practices in early career researchers is invaluable in ensuring the impact of their computational outputs, but software development does not always receive as much attention in student scientific computational training as it deserves. Our experience running doctoral training programs focused on software development for 15 years has shown that this is best achieved through active learning on ambitious group projects targeting real-world problems, mentored by older students on the program. Feedback from students has supported these views, with many students being enthusiastic about learning and applying sustainable software development practices using the approaches we have discussed here and continuing to apply the practices in their work after the conclusion of the course. This teaching pathway has demonstrated a track record of success in open-source software development, with a range of widely used software packages developed by students during their PhDs, which is available at \url{https://github.com/SABS-R3/software-outputs}. For example, students who developed an automated pipeline to build antibody databases during this teaching program have subsequently used this domain-specific experience to publish highly cited antibody structure prediction tools \cite{Abanades2022}, antibody language models \cite{Olsen2022}, and antibody paratope predictors \cite{Chinery2022}. In each case, the incorporation of techniques such as automated workflows, contribution templates and module packaging that they learned during the teaching course has been instrumental in the widespread sharing and reuse of their software outputs.

We hope that other graduate training schemes may consider adopting these strategies in their own institutions, to promote the value of sustainable software development, and the research benefits it can bring.

\section*{Acknowledgments}
We acknowledge the work of Steve Crouch and James Graham from the Software Sustainability Institute, for their helpful instruction on the topic of software engineering in the graduate training courses. We also thank David Augustin for providing helpful comments on the notes which led to this paper. 

\bibliography{references}

\end{document}